%% file: conference-stability-relative.tex
\documentclass{article}
\usepackage[english]{babel}
\usepackage[utf8]{inputenc}
\usepackage{spconf,ifpdf}
\usepackage{cite} 
\usepackage{url}
\usepackage[draft,bookmarks=false]{hyperref} 
\ifpdf
	\usepackage[pdftex]{graphicx}
	\graphicspath{{./figures/}}
\else
	\usepackage[dvips]{graphicx}
	\graphicspath{./figures/}
\fi
\usepackage{color}
\usepackage{pgf, tikz, pgfplots}

\usetikzlibrary{shapes, arrows, automata}
\usepackage{caption} 
\usepackage{subcaption} 
\usepackage{amsmath}
\usepackage{amsfonts, amssymb, amsthm}
\usepackage{mathrsfs}
\usepackage{upgreek}
\usepackage{algorithm,algpseudocode}
\usepackage{enumerate}
\usepackage{multirow}
\usepackage{rotating}
\usepackage{bm}
\usepackage{dsfont}
\input{mySections.tex}

\input{mySymbol.sty}

\newtheorem{assumption}{\hspace{0pt}\bf Assumption}

\newtheorem{proposition}{\hspace{0pt}\bf Proposition}

\newtheorem{theorem}{\hspace{0pt}\bf Theorem}

\newtheorem{definition}{\hspace{0pt}\bf Definition}

\newcommand{\QED}{\hfill\ensuremath{\blacksquare}}

\title{Stability of Neural Networks on Manifolds\\ to Relative Perturbations}
{\name{Zhiyang Wang \qquad Luana Ruiz \qquad Alejandro Ribeiro}
\address{  Department of Electrical and Systems Engineering, University of Pennsylvania, PA 
\thanks{Supported by NSF CCF 1717120, Theorinet Simons. } }
}

\begin{document}
%
\maketitle
%
\begin{abstract}
Graph Neural Networks (GNNs) show impressive performance in many practical scenarios, which can be largely attributed to their stability properties. Empirically, GNNs can scale well on large size graphs, but this is contradicted by the fact that existing stability bounds grow with the number of nodes. Graphs with well-defined limits can be seen as samples from manifolds. Hence, in this paper, we analyze the stability properties of convolutional neural networks on manifolds to understand the stability of GNNs on large graphs. Specifically, we focus on stability to relative perturbations of the Laplace-Beltrami operator. To start, we construct frequency ratio threshold filters which separate the infinite-dimensional spectrum of the Laplace-Beltrami operator. We then prove that manifold neural networks composed of these filters are stable to relative operator perturbations. As a product of this analysis, we observe that manifold neural networks exhibit a trade-off between stability and discriminability. 
Finally, we illustrate our results empirically in a wireless resource allocation scenario {where the transmitter-receiver pairs are assumed to be sampled from a manifold.} 

\end{abstract}
\begin{keywords}
Deep neural networks, manifolds, stability analysis, relative perturbations
\end{keywords}
%


\section{Introduction} \label{sec:intro}

\input{intro}


\section{Preliminary definitions} \label{sec:prelim}

\input{prelims}


\section{Stability of Manifold Neural Networks} \label{sec:filters}

\input{filters_manifold}


\section{Numerical experiments} \label{sec:sims}

\input{simulations}


\section{Conclusions} \label{sec:conclusions}

\input{conclusions}


\bibliographystyle{IEEEbib}
\bibliography{references}

\end{document}

%% file: mySections.tex
\usepackage{needspace}



%% file: intro.tex

Graph Neural Networks (GNNs) are convolutional neural networks architectures where each layer contains a bank of graph convolutional filters followed by a point-wise nonlinearity \cite{gama2019convolutional, zhou2020graph, xu2018powerful}. They have wide applications including but not limited to recommendation systems \cite{fan2019graph}, robot swarms \cite{tolstaya2020learning} and wireless communication networks \cite{wang2020unsupervised, chowdhury2021unfolding}. In these applications, their impressive empirical performance is largely attributed to the invariance and stability properties inherited from convolutions. Akin to the translation equivariance and stability properties of CNNs \cite{mallat2012group}, GNNs have been shown to be permutation equivariant and stable to perturbations of the graph \cite{gama2020stability,zou2020graph}.


Empirically, another important characteristic of GNNs is that they scale well to very large graphs \cite{eisen2020optimal,tolstaya2020learning}. However, this is not reflected in existing stability analyses, where the stability bounds grow with the size of the graph \cite{gama2020stability, zou2020graph}. We posit that tighter bounds can be derived by focusing on graph limits. Leveraging the fact that graphs with arbitrary size and well-defined limits can be seen as discrete samples from a manifold \cite{levie2019transferability,calder2019improved}, in this paper we introduce manifold neural networks (MNNs), and study their stability properties to understand the stability properties of GNNs supported on large graphs.

The MNN is defined as a convolutional architecture on {an embedded manifold in $\reals^N$}, and we are interested in its stability to relative perturbations of the Laplace-Beltrami operator $\ccalL$ (Definition \ref{defn:relative_perturbations}). The Laplace-Beltrami operator is a Laplacian-like operator which is defined locally on the manifold and has a countable real spectrum. The manifold convolution is a pointwise operation on the spectrum of $\ccalL$.
Since relative perturbations of $\ccalL$ result in perturbations of its spectra, the stability of manifold neural networks depends on the stability of its convolutional filters. In this paper, we show that this stability is achievable by Frequency Ratio Threshold (FRT) filters (Definition \ref{def:gamma-filter}), which separate the spectrum into finite groups of eigenvalues. We further show that manifold neural networks composed of FRT filters are stable to relative perturbations of $\ccalL$ (Theorem \ref{thm:stability_nn}), and that there is a trade-off between the stability and the discriminative power of MNNs.

A large number of studies have focused on the stability of GNNs, such as \cite{gama2019stability} and \cite{gama2020stability} which consider absolute and relative graph perturbations respectively. The stability of neural networks on three-dimensional meshes is studied in \cite{kostrikov2018surface}. Prior work on the stability of GNNs in the limit of large graphs includes \cite{ruiz2021graphon1}, which introduces the stability of GNNs to perturbations of dense random graph models called graphons; and \cite{wang2021stability}, which analyzes the stability of MNNs to absolute perturbations of the Laplace-Beltrami operator. In this paper, we extend upon the results of \cite{wang2021stability} by considering relative perturbations, which are a more realistic perturbation model that accounts for the structure of the Laplace-Beltrami operator.
Other related work includes transferability analyses of GNNs considering graphons \cite{ruiz2020graphon,ruiz2021graphon,ruiz2021graphon1}, continuous graph models with tunable sparsity \cite{keriven2020convergence}, and general topological spaces \cite{levie2019transferability}.

The rest of this paper is organized as follows. We introduce the notions of manifold signals and manifold convolution in Section \ref{sec:prelim}. We define FRT filters and prove the stability of manifold neural networks composed of these filters to relative perturbations of the Laplace-Beltrami operator in Section \ref{sec:filters}. We verify our results numerically in a wireless resource allocation scenario in Section \ref{sec:sims}. Concluding remarks are presented in Section \ref{sec:conclusions}.

%% file: prelims.tex
Next we introduce the concepts of a manifold signal, of the Laplace-Beltrami operator and of a manifold convolution. These concepts are necessary to define manifold neural networks and analyze their stability in Section \ref{sec:filters}.


\subsection{Manifolds and manifold signals}

A differentiable $d$-dimensional manifold $\ccalM$ is a topological space where each point $x\in\ccalM$ has a neighborhood that is homeomorphic to a $d$-dimensional Euclidean space which is given by the tangent space $T_x \ccalM$. 
{We consider a simple case of $d$-dimensional embedded submanifold in $\reals^N$.} The collections of scalar functions which map each $x\in\ccalM$ to some real value, and tangent vector functions attaching a tangent vector to each $x\in\ccalM$, are denoted as $L^2(\ccalM)$ and $L^2(T\ccalM)$ respectively, where $T\ccalM$ stands for the disjoint union of all tangent spaces on $\ccalM$. 
We restrict attention to compact and smooth manifolds.   



Manifold signals are defined as data supported on the manifold $\ccalM$ and written as scalar functions $f\in L^2(\ccalM)$ attaching some real value $f(x)$ to each point $x\in\ccalM$. 
For these signals, differentiation is defined as the application of an operator $\nabla: L^2(\ccalM)\rightarrow L^2(T\ccalM)$ called \emph{intrinsic gradient} \cite{bronstein2017geometric}. Given a signal $f$, $\nabla f(x)$ indicates the fastest changing direction of a function at the point $x$, which is represented by a vector on the tangent space of $x$. The adjoint of the intrinsic gradient operator is the \emph{intrinsic divergence}, denoted $\text{div}: L^2(T\ccalM)\rightarrow L^2(\ccalM)$. By composing these two operators, the Laplace-Beltrami operator is defined as
\begin{equation}\label{eqn:Laplacian}
    \ccalL f=-\text{div}(\nabla f).
\end{equation}
Similarly to Laplacian operators in Euclidean domains, the Laplace-Beltrami operator measures the total variation of a function by quantifying the difference between the instantaneous function value at a given point, and the local average of the function around that point.

The Laplace-Beltrami operator $\ccalL$ is a self-adjoint and positive-semidefinite operator by definition. Therefore, it possesses a discrete spectrum $\{\lambda_i,\bm\phi_i\}_{i\in\naturals^+}$, where $\lambda_i$ are real positive eigenvalues and $\bm\phi_i$ are the corresponding eigenfunctions. Explicitly, we can write $\ccalL$ as
\begin{equation} \label{eqn:Laplacian-spectrum}
    \ccalL f=\sum_{i=1}^\infty \lambda_i\langle f, \bm\phi_i \rangle \bm\phi_i
\end{equation}
where the eigenvalues are ordered in increasing order as $0<\lambda_1\leq \lambda_2\leq\lambda_3\leq \hdots$ and, according to Weyl's law \cite{arendt2009weyl}, grow as $i^{2/d}$ where $d$ is the manifold dimension. The eigenfunctions form an orthonormal basis of $L^2(\ccalM)$ which is also intrinsic because of the intrinsic construction of $\ccalL$. 
As such, a square-integrable function $f\in L^2(\ccalM)$ can be represented on this basis as $f=\sum_{i=1}^\infty \langle f, \bm\phi_i \rangle \bm\phi_i$. The $\lambda_i$ are interpreted as  \textit{manifold frequencies} and the $\phi_i$ as  \textit{manifold oscillation modes}.


\subsection{Manifold convolutions and manifold neural networks}
The Laplace-Beltrami operator spectrum \eqref{eqn:Laplacian-spectrum} allows defining the spectral convolution of a manifold signal. 
Namely, the spectral convolutional filter is defined as
\begin{equation}\label{eqn:operator}
\bbh(\ccalL) f:=\sum_{i=1}^\infty \sum_{k=0}^{K-1} h_k \lambda_i^k \langle f,\bm\phi_i \rangle \bm\phi_i,
\end{equation}
where $h_0, \ldots, h_{K-1}$ are the filter coefficients.
Projecting \eqref{eqn:operator} onto the Laplace-Beltrami operator eigenbasis, we see that the spectral response of the manifold convolution is given by the function $h(\lambda) = \sum_{k=0}^{K-1} h_k \lambda_k$ evaluated at the eigenvalues $\lambda_i$.
This indicates that the frequency response of the manifold convolution is decided solely by the coefficients $h_k$---or, equivalently, the filter function $h(\lambda)$---and by the eigenvalues of the Laplace-Beltrami operator. Hence, we could implement the same manifold filter on a new manifold $\ccalM'$ by simply replacing the Laplace-Beltrami operator $\ccalL$ with the new operator $\ccalL'$, in which case the output of the convolution \eqref{eqn:operator} would be determined by the spectrum of $\ccalL'$.

With the manifold convolution operation defined as in \eqref{eqn:operator}, we define the Manifold Neural Networks (MNNs) as a cascade of $L$ layers where each layer contains a bank of manifold convolutional filters followed by a nonlinear activation function. Letting $\sigma$ denote the activation function, the $l$-th layer of a $L$-layer CNN on manifold $\ccalM$ is written as
\begin{equation}\label{eqn:mnn}
f_l^p(x) = \sigma\left( \sum_{q=1}^{F_{l-1}} \bbh_l^{pq}(\ccalL) f_{l-1}^q(x)\right), \quad l=1,2\hdots,L,
\end{equation}
where the $\bbh_l^{pq}(\ccalL)$ are filters mapping the $q$-th feature of the $l-1$-th layer to the $p$-th feature of the $l$-th layer for $1\leq q\leq F_{l-1}$ and $1\leq p\leq F_{l}$. The output features of the last layer, i.e., the output of the neural network, are $f_L^p$ with $1 \leq p \leq F_L$. 
The input features of the first layer, i.e., the input data, are $f^q$ with $1\leq q\leq F_0$. Alternatively, we may write this manifold neural network as a map $ \bbPhi(\bbH,\ccalL, f)$ where the tensor $\bbH$ gathers all the learnable parameters from all layers. 


%% file: filters_manifold.tex

To establish the stability properties of MNNs
, we start by looking at the effect relative perturbations have on the manifold convolutional filters that compose their layers. Relative perturbations of the Laplace-Beltrami operator are defined as follows.


\begin{definition}[Relative perturbations] \label{defn:relative_perturbations}
Let $\ccalL$ be the Laplace-Beltrami operator of an embedded manifold $\ccalM$. A relative perturbation of $\ccalL$ is defined as
\begin{equation}\label{eqn:perturb}
\ccalL'=\ccalL+\bbE\ccalL,
\end{equation}
where the relative perturbation operator $\bbE$ is symmetric.
\end{definition}

The relative perturbation model in Definition \ref{defn:relative_perturbations} describes perturbations that scale the Laplace-Beltrami operator while preserving its symmetries. This makes for a more realistic perturbation model than (absolute) additive perturbations, because it respects the structure of the original Laplace-Beltrami operator. Thinking of the manifold as a continuous graph limit, relative perturbations can be seen as perturbing each edge proportionally to its edge weight.

\subsection{Frequency ratio threshold (FRT) filters}

From the eigendecomposition of the Laplace-Beltrami operator \eqref{eqn:Laplacian}, it is clear that a perturbation of $\ccalL$ will result in some sort of perturbation of its spectrum. Since the spectral convolution \eqref{eqn:operator} depends directly on $\lambda_i$ through its frequency response $h(\lambda_i)$, each individual eigenvalue can affect the stability of the output signal. Hence, we have to analyze the effect of each eigenvalue perturbation individually. What makes the problem challenging in the manifold setting is that, though countable, the spectrum of Laplace-Beltrami operator $\ccalL$ is infinite-dimensional. However, it is possible to show that the eigenvalues accumulate in certain parts the spectrum by Weyl's law \cite{arendt2009weyl}. This result is stated in Proposition \ref{prop:finite_num}.


\begin{proposition} \label{prop:finite_num}
Let $\ccalM$ be a $d$-dimensional embedded manifold in $\reals^N$ with Laplace-Beltrami operator $\ccalL$, and let $\{\lambda_k\}_{k=1}^\infty$ denote the eigenvalues of $\ccalL$. Let $C_1$ denote an arbitrary constant. For any $\gamma > 0$, there exists $N_1$ given by
\begin{equation}
    N_1=\lceil (C_1 (\gamma+1)^{d/2}-1)^{-1} \rceil
\end{equation}
such that, for all $k>N_1$, it holds that $$\lambda_{k+1}-\lambda_k\leq \gamma\lambda_k.$$
\end{proposition}
\begin{proof}
This is a direct consequence of Weyl's law \cite{arendt2009weyl}.
\end{proof}

Proposition \ref{prop:finite_num} implies that, for large enough eigenvalues, the distance between two consecutive eigenvalues is at most $\gamma$ times the smallest eigenvalue, where $\gamma$ is positive but can be as small as desired. Thus, we can group eigenvalues whose differences to neighboring eigenvalues are no more than a small scaling (measured by $\gamma$) of their own magnitude. This allows partitioning the spectrum into a $\gamma$-separated spectrum where the ratio between the largest neighboring eigenvalue to $\lambda_k$, i.e., $\lambda_{k+1}$, and $\lambda_k$ itself is smaller than $1+\gamma$. This is described in Definition \ref{def:gamma-spectrum}. In Definition \ref{def:gamma-filter}, we further define Frequency Ratio Threshold (FRT) filters, which are filters that can separate the spectrum in this way.


\begin{definition}[$\gamma$-separated spectrum]\label{def:gamma-spectrum}
The $\gamma$-separated spectrum of a Laplace-Beltrami operator $\ccalL$ is defined as the partition $\Lambda_1(\gamma)\cup\hdots \cup\Lambda_M(\gamma)$ such that, for $l\neq l$, all $\lambda_i\in\Lambda_k(\gamma)$ and $\lambda_j\in\Lambda_l(\gamma)$, satisfy
\begin{equation}
\label{eqn:frt-spectrum}
\left|\frac{\lambda_i}{\lambda_j}-1 \right|>\gamma.
\end{equation}
\end{definition}

\begin{definition}[$\gamma$-FRT filter]\label{def:gamma-filter}
The $\gamma$-frequency ratio threshold ($\gamma$-FRT) filter is a manifold filter $\bbh(\ccalL)$ whose frequency response satisfies
\begin{equation}\label{eqn:frt-filter}
       |h(\lambda_i)-h(\lambda_j)|\leq \Delta_k,\text{ for all } \lambda_i,\lambda_j\in\Lambda_k(\gamma)
\end{equation}
with $\Delta_k\leq \Delta$ for $k=1,2\hdots,M$.
\end{definition}

\begin{figure}[t]
    \centering
    \includegraphics[width=0.4\textwidth]{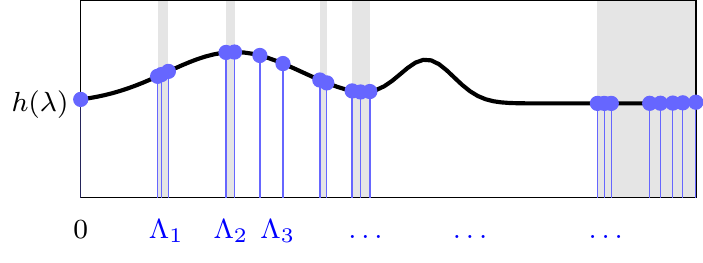}  
    \caption{A $\gamma$-FRT filter that separates the spectrum of the Laplacian operator. The $x$-axis stands for the spectrum with each sample representing an eigenvalue. The gray shade shows the grouping of the eigenvalues.}
    \label{fig:filter-response}
\end{figure}

In the $\gamma$-separated spectrum, eigenvalues $\lambda_i\in \Lambda_k$ and $\lambda_j\in\Lambda_l$ in different groups are at least $\gamma\min(\lambda_i,\lambda_j)$ apart. In other words, the spectrum separation realized by a $\gamma$-FRT filter is such that eigenvalues are separated by \textit{relative} eigenvalue distances.
Note that the $\gamma$-FRT filter achieves spectrum separation by treating eigenvalues $\lambda_i,\lambda_j\in\Lambda_k(\gamma)$ similarly, i.e., by giving them spectral responses whose difference is bounded. However, the spectral responses can vary freely for eigenvalues in different groups as shown in Figure \ref{fig:filter-response}.

\subsection{Manifold Neural Network Stability}

In order to prove stability of MNNs, we need the following two assumptions.

\begin{assumption}\label{ass:filter_function}
The filter function $h:\reals\rightarrow\reals$ is $B$- integral Lipschitz continuous and non-amplifying, i.e.,
\begin{equation}\label{eqn:filter_function}
    |h(a)-h(b)|\leq \frac{B|a-b|}{(a+b)/2},\quad |h(a)|< 1\quad \text{for all }a,b.
\end{equation}
\end{assumption}

\begin{assumption}[Normalized Lipschitz activation functions]\label{ass:activation}
 The activation function $\sigma$ is normalized Lipschitz continuous, i.e., $|\sigma(a)-\sigma(b)|\leq |a-b|$, with $\sigma(0)=0$.
\end{assumption}

The integral Lipschitz condition can be understood as a Lipschitz continuity condition with constant $2B/(a+b)$. When $a$ and $b$ are close, this condition can be approximated by $|ah'(a)|\leq B$ as illustrated in Figure \ref{fig:filter-response}. Most common activation functions (e.g. ReLu, modulus and sigmoid) satisfy Assumption \ref{ass:activation}. 
Under these assumptions, MNNs with $\gamma$-FRT manifold filters are thus stable to relative perturbations (Definition \ref{defn:relative_perturbations}) as stated in Theorem \ref{thm:stability_nn}.

\begin{theorem}[Neural network stability]\label{thm:stability_nn}
 Let $\ccalM$ be a manifold with Laplace-Beltrami operator $\ccalL$. Let $f$ be a manifold signal and $\bm\Phi(\bbH,\ccalL,f)$ an $L$-layer manifold neural network on $\ccalM$ \eqref{eqn:mnn} with $F_0=F_L=1$ input and output features and $F_l=F$ features per layer for $l=1,2,\hdots,L-1$. Let the filters $\bbh^{pq}_l(\ccalL)$ be $\gamma$-FRT [cf. Definition \ref{def:gamma-filter}] with $\Delta=\frac{\pi\epsilon}{\gamma-\epsilon+\gamma\epsilon}$ and $B$-integral Lipschitz. 
Let $\ccalL'=\ccalL + \bbE\ccalL$ be a relative perturbation of the Laplace-Beltrami operator $\ccalL$ [cf. Definition \ref{defn:relative_perturbations}] where $\|\bbE\| = \epsilon \leq \gamma$. 
 Under Assumptions \ref{ass:filter_function} and \ref{ass:activation}, it holds that
 \begin{align}\label{eqn:stability_nn}
 \begin{split}
    \|\bm\Phi(\bbH,\ccalL,f)&-\bm\Phi(\bbH,\ccalL',f)\| \\
    &\leq LF^{L-1}\left( \frac{2M\pi}{\gamma-\epsilon+\gamma\epsilon}+\frac{2B}{2-\epsilon} \right)\epsilon \|f\|.
\end{split}
 \end{align}
where $M$ is the number of the partitions [cf. Definition \ref{def:gamma-spectrum}]. 
 \end{theorem}

When $\epsilon$ is sufficiently small ($\epsilon \ll \min(\gamma,2)$,
the denominators on the right hand side of \eqref{eqn:stability_nn} are approximately equal to $\gamma$ and $2$ respectively. Thus, MNNs with $\gamma$-FRT integral Lipschitz filters are stable to relative perturbations of the Laplace-Beltrami operator. 
The frequency ratio threshold $\gamma$ affects stability directly (by appearing in the bound in Theorem \ref{thm:stability_nn}) and indirectly through the partition size $M$. With a larger $\gamma$, more eigenvalues will be in the same group, thus decreasing $M$ and improving stability. A smaller integral Lipschitz constant $B$ also increases stability. However, small $B$ and large $\gamma$ make for smoother filters which in turn lead to a less discriminative neural network. Therefore, MNNs with integral Lipschitz $\gamma$-FRT filters exhibit a trade-off between discriminality and stability. However, this lack of discriminality can be lifted by the pointwise nonlinearity. It can spread the information in a signal throughout the whole spectrum by creating frequency responses in frequencies that do not have responses before. The stability bound also scales with the size of the neural network.



%% file: simulations.tex
MNNs can be seen as limits of GNNs. Hence, they cannot be implemented in practice, but we can illustrate our results numerically using GNNs \cite{levie2019transferability}. Specifically, we consider a wireless resource allocation scenario. We construct a wireless adhoc network by dropping $n=50$ nodes randomly over a range of $[-50m,50m]^2$. The fading link states can be represented by a matrix $\bbS(t)$, each element $[\bbS(t)]_{ij}:=s_{ij}(t)$ of which represents the channel condition between node $i$ and node $j$. When considering the large-scale pathloss gain and a random fast fading gain, the link state can be written as  $s_{ij}=\log( d_{ij}^{-2.2} h^f)$, where $d_{ij}$is the  distance between node $i$ and $j$, while $h^f\sim  \text{Rayleigh}(2)$ is the random fading. We consider the power allocation problem among $n$ nodes over an AWGN channel. The goal is to maximize the sum-of-rate capacity under a total power budget $P_{max}$ with $\bbp(\bbS)=[p_1,p_2,\hdots,p_n]$ denoting the power allocated to each node under channel condition $\bbS$ and the channel rate of node $i$ represented as $r_i$. The problem can be formulated as
\begin{align}
\label{eqn:prob_sim}
r^*&=\max_{\bbp(\bbS)} \sum_{i=1}^n r_i\\
   s.t.\quad \nonumber &r_i=\mathbb{E}\left[  \log\left(1+\frac{|h_{ii}|^2 p_i(\bbS)}{1+ \sum\limits_{j\neq i} |s_{ij}|^2 p_j(\bbS)}\right) \right],\\
   \nonumber & \mathbb{E}[\bm{1}^T\bbp]\leq P_{max},\quad p_i(\bbS)\in \{0,p_0\}.
\end{align}

\begin{figure}[h]
    \centering
    \includegraphics[width=0.35\textwidth]{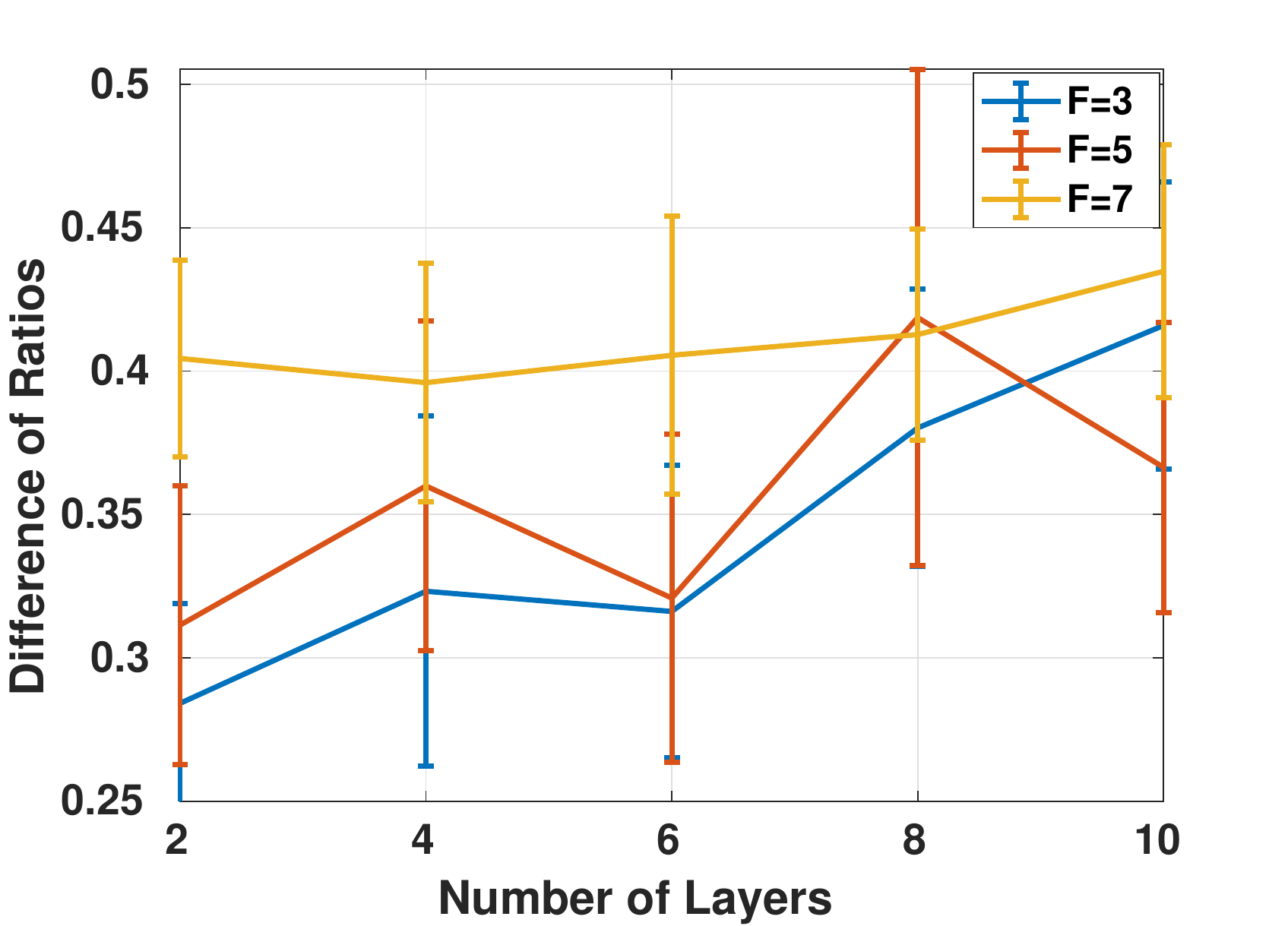}
    \caption{Sum-of-rate ratio differences on the test between the original wireless network setting and the perturbed one.}
    \label{fig:sim_stability}
\end{figure}

The nodes and links in the wireless setting can be seen as graph nodes and edges. By formulating the adjacency matrix $\bbS$ into a graph Laplacian, the problem can be solved with a GNN containing the proposed $\gamma$-FRT filters. After trained for $4000$ iterations, the GNN can achieve the optimal performance. In practice, the nodes are often deployed in dynamic environments which would cause perturbations to the underlying Laplacian matrix. To model this, we add a log-normal matrix to scale the original channel matrix $\bbS$. With the same trained GNN employed, we measure the stability by the difference of ratio of final sum-of-rate to a baseline sum-of-rate. We can observe from Figure \ref{fig:sim_stability} that the difference increases with the number of layers and the number of filters per layer in the constructed GNN, but is overall small. This justifies the result that we have proposed in Theorem \ref{thm:stability_nn}.

%% file: conclusions.tex
In this paper, we introduced manifold convolutional filters and manifold neural networks. Additionally, we defined $\gamma$-frequency ratio threshold filters that separate the infinite-dimensional spectrum of the Laplace-Beltrami operator into finite groups. By treating eigenvalues that are relatively close to each other similarly, manifold neural networks composed of these filters were shown to be stable under relative perturbations of the Laplace-Beltrami operator. Furthermore, they exhibited a trade-off between stability and discriminability. These results were verified empirically on a power allocation problem in wireless adhoc networks assumed sampled from a manifold.

%% file: conference-stability-relative.bbl
\begin{thebibliography}{10}

\bibitem{gama2019convolutional}
Fernando Gama, Antonio~G Marques, Geert Leus, and Alejandro Ribeiro,
\newblock ``Convolutional neural network architectures for signals supported on
  graphs,''
\newblock {\em IEEE Transactions on Signal Processing}, vol. 67, no. 4, pp.
  1034--1049, 2019.

\bibitem{zhou2020graph}
Jie Zhou, Ganqu Cui, Shengding Hu, Zhengyan Zhang, Cheng Yang, Zhiyuan Liu,
  Lifeng Wang, Changcheng Li, and Maosong Sun,
\newblock ``Graph neural networks: A review of methods and applications,''
\newblock {\em AI Open}, vol. 1, pp. 57--81, 2020.

\bibitem{xu2018powerful}
Keyulu Xu, Weihua Hu, Jure Leskovec, and Stefanie Jegelka,
\newblock ``How powerful are graph neural networks?,''
\newblock {\em arXiv preprint arXiv:1810.00826}, 2018.

\bibitem{fan2019graph}
Wenqi Fan, Yao Ma, Qing Li, Yuan He, Eric Zhao, Jiliang Tang, and Dawei Yin,
\newblock ``Graph neural networks for social recommendation,''
\newblock in {\em The World Wide Web Conference}, 2019, pp. 417--426.

\bibitem{tolstaya2020learning}
Ekaterina Tolstaya, Fernando Gama, James Paulos, George Pappas, Vijay Kumar,
  and Alejandro Ribeiro,
\newblock ``Learning decentralized controllers for robot swarms with graph
  neural networks,''
\newblock in {\em Conference on robot learning}. PMLR, 2020, pp. 671--682.

\bibitem{wang2020unsupervised}
Zhiyang Wang, Mark Eisen, and Alejandro Ribeiro,
\newblock ``Unsupervised learning for asynchronous resource allocation in
  ad-hoc wireless networks,''
\newblock {\em arXiv preprint arXiv:2011.02644}, 2020.

\bibitem{chowdhury2021unfolding}
Arindam Chowdhury, Gunjan Verma, Chirag Rao, Ananthram Swami, and Santiago
  Segarra,
\newblock ``Unfolding wmmse using graph neural networks for efficient power
  allocation,''
\newblock {\em IEEE Transactions on Wireless Communications}, 2021.

\bibitem{mallat2012group}
St{\'e}phane Mallat,
\newblock ``Group invariant scattering,''
\newblock {\em Communications on Pure and Applied Mathematics}, vol. 65, no.
  10, pp. 1331--1398, 2012.

\bibitem{gama2020stability}
Fernando Gama, Joan Bruna, and Alejandro Ribeiro,
\newblock ``Stability properties of graph neural networks,''
\newblock {\em IEEE Transactions on Signal Processing}, vol. 68, pp.
  5680--5695, 2020.

\bibitem{zou2020graph}
Dongmian Zou and Gilad Lerman,
\newblock ``Graph convolutional neural networks via scattering,''
\newblock {\em Applied and Computational Harmonic Analysis}, vol. 49, no. 3,
  pp. 1046--1074, 2020.

\bibitem{eisen2020optimal}
Mark Eisen and Alejandro Ribeiro,
\newblock ``Optimal wireless resource allocation with random edge graph neural
  networks,''
\newblock {\em IEEE Transactions on Signal Processing}, vol. 68, pp.
  2977--2991, 2020.

\bibitem{levie2019transferability}
Ron Levie, Michael~M Bronstein, and Gitta Kutyniok,
\newblock ``Transferability of spectral graph convolutional neural networks,''
\newblock {\em arXiv preprint arXiv:1907.12972}, 2019.

\bibitem{calder2019improved}
Jeff Calder and Nicolas~Garcia Trillos,
\newblock ``Improved spectral convergence rates for graph laplacians on
  epsilon-graphs and k-nn graphs,''
\newblock {\em arXiv preprint arXiv:1910.13476}, 2019.

\bibitem{gama2019stability}
Fernando Gama, Joan Bruna, and Alejandro Ribeiro,
\newblock ``Stability of graph scattering transforms,''
\newblock {\em arXiv preprint arXiv:1906.04784}, 2019.

\bibitem{kostrikov2018surface}
Ilya Kostrikov, Zhongshi Jiang, Daniele Panozzo, Denis Zorin, and Joan Bruna,
\newblock ``Surface networks,''
\newblock in {\em Proceedings of the IEEE Conference on Computer Vision and
  Pattern Recognition}, 2018, pp. 2540--2548.

\bibitem{ruiz2021graphon1}
Luana Ruiz, Zhiyang Wang, and Alejandro Ribeiro,
\newblock ``Graphon and graph neural network stability,''
\newblock in {\em ICASSP 2021-2021 IEEE International Conference on Acoustics,
  Speech and Signal Processing (ICASSP)}. IEEE, 2021, pp. 5255--5259.

\bibitem{wang2021stability}
Zhiyang Wang, Luana Ruiz, and Alejandro Ribeiro,
\newblock ``Stability of neural networks on riemannian manifolds,''
\newblock {\em arXiv preprint arXiv:2103.02663}, 2021.

\bibitem{ruiz2020graphon}
Luana Ruiz, Luiz Chamon, and Alejandro Ribeiro,
\newblock ``Graphon neural networks and the transferability of graph neural
  networks,''
\newblock {\em Advances in Neural Information Processing Systems}, vol. 33,
  2020.

\bibitem{ruiz2021graphon}
Luana Ruiz, Luiz~FO Chamon, and Alejandro Ribeiro,
\newblock ``Graphon signal processing,''
\newblock {\em IEEE Transactions on Signal Processing}, vol. 69, pp.
  4961--4976, 2021.

\bibitem{keriven2020convergence}
Nicolas Keriven, Alberto Bietti, and Samuel Vaiter,
\newblock ``Convergence and stability of graph convolutional networks on large
  random graphs,''
\newblock {\em arXiv preprint arXiv:2006.01868}, 2020.

\bibitem{bronstein2017geometric}
Michael~M Bronstein, Joan Bruna, Yann LeCun, Arthur Szlam, and Pierre
  Vandergheynst,
\newblock ``Geometric deep learning: going beyond euclidean data,''
\newblock {\em IEEE Signal Processing Magazine}, vol. 34, no. 4, pp. 18--42,
  2017.

\bibitem{arendt2009weyl}
Wolfgang Arendt, Robin Nittka, Wolfgang Peter, and Frank Steiner,
\newblock ``Weyl’s law: Spectral properties of the laplacian in mathematics
  and physics,''
\newblock {\em Mathematical analysis of evolution, information, and
  complexity}, pp. 1--71, 2009.

\end{thebibliography}
